\documentclass[conference,keeplastbox]{IEEEtran} 
\usepackage{amssymb,amsfonts,amsmath}
\usepackage{dsfont}
\usepackage{balance}
\usepackage{cite}
\usepackage{flushend}

\IEEEoverridecommandlockouts

\ifCLASSINFOpdf
   \usepackage[pdftex]{graphicx}
   \DeclareGraphicsExtensions{.pdf,.jpeg,.png}
\else
\fi
\def\(({\left(}
\def\)){\right)}
\def\[[{\left[}
\def\]]{\right]}

\newcommand{\be}{\begin{equation}}
\newcommand{\ee}{\end{equation}}
\newcommand{\bea}{\begin{eqnarray}}
\newcommand{\eea}{\end{eqnarray}}

\newcommand{\BEAS}{\begin{eqnarray*}}
\newcommand{\EEAS}{\end{eqnarray*}}
\newcommand{\BEA}{\begin{eqnarray}}
\newcommand{\EEA}{\end{eqnarray}}

\usepackage[utf8]{inputenc}
\usepackage{bm}
\usepackage{xcolor}
\usepackage{algpseudocode}
\graphicspath{{./figures/}}

\begin{document}
%
\title{Multi-Layer Generalized Linear Estimation}

\author{
    \IEEEauthorblockN{Andre Manoel}
        \IEEEauthorblockA{Neurospin, CEA\\
                          Universit{\'e} Paris-Saclay}
\and
\IEEEauthorblockN{Florent Krzakala}
        \IEEEauthorblockA{LPS ENS, CNRS \\
                          PSL, UPMC \& Sorbonne Univ.}
\and
 \IEEEauthorblockN{Marc M\'ezard}
        \IEEEauthorblockA{Ecole Normale Sup\'erieure\\
          PSL Research University}
\and
        \IEEEauthorblockN{Lenka Zdeborov{\'a}}
        \IEEEauthorblockA{IPhT, CNRS, CEA\\
                          Universit{\'e} Paris-Saclay}
}



\maketitle

\begin{abstract}
  We consider the problem of reconstructing a signal from
  multi-layered (possibly) non-linear measurements. Using non-rigorous but
  standard methods from statistical physics we present the
  Multi-Layer Approximate Message Passing (ML-AMP) algorithm for
  computing marginal probabilities of the corresponding estimation
  problem and derive the associated state evolution equations to
  analyze its performance. We also give the expression of the
  asymptotic free energy and the minimal information-theoretically
  achievable reconstruction error. Finally, we present some
  applications of this measurement model for compressed sensing and
  perceptron learning with structured matrices/patterns, and for a
  simple model of estimation of latent variables in an auto-encoder.
\end{abstract}

\IEEEpeerreviewmaketitle
In many natural and engineered systems, the interactions between sets of
variables in different subsystems involve multiple layers of
interdependencies.
This is for instance the case in the neural networks
developed in deep learning \cite{lecun2015deep}, the
hierarchical models used in statistical inference
\cite{ranganath_hierarchical_2015}, and the multiplex networks
considered in complex systems \cite{mucha2010community}. It is
therefore fundamental to generalize our theoretical and algorithmic
tools to deal with these \emph{multi-layer} setups. Our goal in this paper
is to develop such a generalization of the cavity/replica approach
that originated in statistical physics \cite{mezard2009information}
and that has been shown to be quite successful for studying
generalized linear estimation with randomly chosen mixing, leading in
particular to the computation of the mutual information (or
equivalently free energy) and minimum achievable mean-squared error
for CDMA and compressed sensing
\cite{tanaka2002statistical,wu2012optimal,barbier2016mutual,reeves2016replica}.
This methodology is also closely related to the
approximate message passing (AMP) algorithm, originally known in physics as
Thouless-Anderson-Palmer (TAP) equations
\cite{thouless1977solution,mezard1989space,1742-6596-95-1-012001,donoho_message-passing_2009,rangan_generalized_2011,krzakala_probabilistic_2012}.

We present in section \ref{sec:problem} a multi-layer generalized linear
measurement (ML-GLM) model with random weights at each layer and
consider the Bayesian inference of the signal measured by the
ML-GLM. We derive AMP for ML-GLM and, using non-rigorous but standard
methods from statistical physics, analyze its behavior by deriving the
associated state evolution. We also present the expression for the
associated free energy (or mutual information) and the optimal
information-theoretically mean squared error
(MMSE). We compare the MMSE with the MSE achieved by AMP and describe
the associated phase transitions.


\section{Problem statement}
\label{sec:problem} 
{\bf ML-GLM model:} Consider $L$ known matrices
$W^{(1)}$, $W^{(2)}$, $\dots$, $W^{(L)}$ of dimension
$W^{(\ell)} \in \mathbb{R}^{n_{\ell - 1} \times n_{\ell}}$. A control
parameter which will be important is the ratio of the number of rows to
columns in each of these matrices,
$\alpha_\ell \!=\! n_{\ell - 1} / n_{\ell}$. The components of each of these
matrices are drawn independently at random, from a probability
distribution $P_{W^{(\ell)}}$ having zero mean and variance $1/n_\ell$.
We consider a signal $\bm{x}\! \in\! \mathbb{R}^{n_L}$ with elements $x_i$, $i=1,\dots,n_L$ sampled independently from a
distribution $P_X (x_i)$.  We then collect $n_0$ observations
$\bm{y} \in \mathbb{R}^{n_0}$ of the signal $\bm{x}$ as
\begin{equation}
    \bm{y} = f^{(1)}_{\xi^1} (W^{(1)} f^{(2)}_{\xi^2} (W^{(2)} \cdots f^{(L)}_{\xi^L}
        (W^{(L)} \bm{x}))),
\label{problem:def}
\end{equation}
where the so-called {\it activation functions} $f^{(\ell)}_{\xi^\ell}$,
$\ell=1,\dots,L$, are applied element-wise. These functions can be
deterministic or stochastic and are, in general, non-linear. Assuming
$f^{(\ell)}_{\xi}(z)$ depends on a variable $z$ and some noise $\xi$
distributed with $P^{(\ell)} (\xi)$, we can define the probability distribution
of the output of the function
$h= f^{(\ell)}_\xi(z)$ as 
\begin{equation}
    P_\text{out}^{(\ell)}(h | z) = \int {\rm d}\xi \, P^{(\ell)} (\xi) \,
       \delta(h - f^{(\ell)}_{\xi}(z)) \, ,
\end{equation}
where $\delta(.)$ is the Dirac function. $P_\text{out}$ is then
interpreted as a noisy channel through which the variable $z$ is
observed, $h$ being the observation. With the above definition of $P_\text{out}$ we can rewrite
eq.~(\ref{problem:def}) in an equivalent form introducing {\it hidden} auxiliary
variables $h^{(\ell)} \in \mathbb{R}^{n_\ell}$ for $\ell=1,\dots,L-1$ as 
        \begin{align}
        y_\mu &\sim P^{(1)}_\text{out} \Big({y}_\mu \Big| \textstyle{\sum_{i=1}^{n_1}}
        W^{(1)}_{\mu i} h^{(1)}_i  \Big),   \label{eq:multilayer} \\
        {h}^{(1)}_\mu & \sim P^{(2)}_\text{out}\Big( {h}^{(1)}_\mu \Big| \textstyle{\sum_{i=1}^{n_2}}
        W^{(2)}_{\mu i} h^{(2)}_i  \Big),  \nonumber\\
        & \vdots \nonumber \nonumber\\
        {h}^{(L-1)}_\mu & \sim P^{(L)}_\text{out} \Big( {h}^{(L-1)}_\mu \Big| \textstyle{\sum_{i=1}^{n_L}}
        W^{(L)}_{\mu i} x_i  \Big), \nonumber \\
        {x}_\mu &\sim P_X ( {x}_\mu), \nonumber
        \end{align}
The inference problem of interest in this paper is the
MMSE estimation of the signal
$\bm{x}$ from the knowledge of the observation $\bm{y}$ and the
matrices $W^{(\ell)}$, for all $\ell=1,\dots,L$. This inference is done
through the computation of marginals of the corresponding posterior distribution
$P(\bm{x}|\bm{y})$.

Using the Bayes theorem and the above definition
of the hidden variables $h^{(\ell)} \in \mathbb{R}^{n_l}$, the posterior is written as 
{\small
\begin{eqnarray}
&& P(\bm{x}|\bm{y}) = \frac{1}{{Z}(\bm{y})} \prod_{\mu = 1}^{n_L} P_X(x_\mu) \int 
  \prod_{\ell=1}^{L-1} \prod_{\mu =1}^{n_\ell} {\rm d}h_\mu^{(\ell)} \,
   \nonumber \\ && \kern-2em
  \prod_{\mu=1}^{n_1}  P_\text{out}^{(1)}\Big (y_\mu \Big|  \sum_{i=1}^{n_1}
   W_{\mu i}^{(1)} h_i^{(1)}\Big) 
  \prod_{\mu=1}^{n_{L}}  P_\text{out}^{(L)}\Big (h^{(L-1)}_\mu \Big|  \sum_{i=1}^{n_{L}} W_{\mu i}^{(L)}  x_i \Big)  
  \nonumber \\ && \kern5em 
   \prod_{\ell=2}^{L-1} \prod_{\mu=1}^{n_\ell}  P_\text{out}^{(\ell)}\Big(h^{(\ell-1)}_\mu \Big|  \sum_{i=1}^{n_\ell} W_{\mu i}^{(\ell)}  h_i^{(\ell)}\Big).
   \label{posterior}
\end{eqnarray}
}%
We focus here on the ``Bayes-optimal inference'' where the generative model and all its
parameters are known, i.e. the only unknown variables are $\bm{x}$ and the
$\bm{h}^{(\ell)}$, $\ell = 1, \dots, L-1$. In this case, in order to minimize the expected mean-squared
error between the ground truth value of $\bm{x}$ and its estimator $\bm{\hat
    x}$ one needs to compute averages of the marginals of the posterior
distribution. 

As usual, computing the marginals of the posterior (\ref{posterior}) is
in general intractable. In this paper we develop an analysis of the
posterior and its marginals that is asymptotically exact, in the sense that, 
with high probability, the estimated marginal probabilities differ
from the true ones by a factor that goes to zero in the
``thermodynamic'' limit $n_\ell\to \infty$
for every $\ell=0,\dots,L$ (at fixed ratios $\alpha_\ell
=n_{\ell-1}/n_\ell=O(1)$). 
Our analysis is based on an AMP-type
algorithm and an analysis of the corresponding
Bethe/replica free energy.

\begin{figure}[t!]
    \centering
    \includegraphics[width=0.45\textwidth]{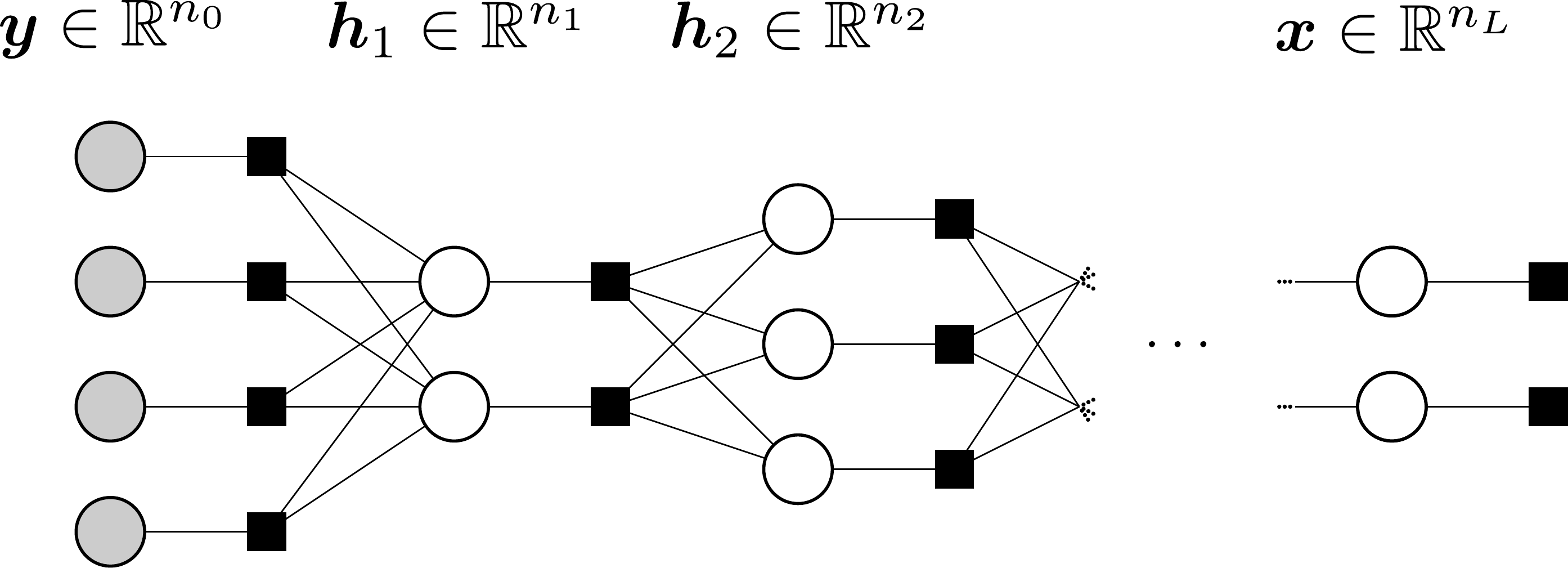}
    \caption{Factor graph of the multi-layer generalized linear
        estimation problem (\ref{posterior}). Shaded circles correspond to
        observations $\bm{y}$, empty circles to hidden variables $\bm{h}$
        and the signal $\bm{x}$ to be inferred. Squares represent the
        activation functions relating the variables via
        eqs.~(\ref{eq:multilayer}). }
    \label{fig:scheme2}\vspace{-3mm}
\end{figure}

{\bf Context:} The ML-GLM model considered in this paper has a range
of applications and is related to other models considered in the
literature. It is similar to the deep exponential families of
\cite{ranganath_hierarchical_2015,ranganath_deep_2015} with fixed
known random weights. It can also be seen as the decoder-side
of an autoencoder neural network with fixed known
weights (corresponding to a randomly generated hierarchy of features),
the goal being to infer the vector of latent variables that is the
closest to some ground-truth values used to generate the data. One
of the main assets of the considered ML-GLM is that the activation functions in
each of the layers can be non-linear, as is the case in deep neural
networks, and our analysis handles these non-linearities. When the intermediate layers are linear, the ML-GLM can be
interpreted as a model for compressed sensing with a structured
measurement matrix obtained as a product of random
matrices. Similarly, when the external layer has a threshold
activation function and all the other activation functions are linear,
the ML-GLM can be seen as a single layer perceptron that aims to
classify correlated patterns obtained by products of random matrices.

We anticipate that the algorithm and theory developed in this paper
will find applications to other learning problems. A natural direction of
future work is to prove both the state evolution of the algorithm and
the Bethe free energy for the ML-GLM rigorously, perhaps along the lines of
\cite{bayati2011dynamics,barbier2016mutual,reeves2016replica}.

\section{Algorithm and analysis}

{\bf ML-AMP:} 
We consider a probability
distribution $P(\bm{x}, \{\bm{h}^{(\ell)}\}_{\ell=1}^{L-1}|\bm{y})$
defined as the posterior distribution (\ref{posterior}) without the
integral over the auxiliary variables
$\{\bm{h}^{(\ell)}\}_{\ell=1}^{L-1}$, and represented by the graphical model depicted in
Fig.~\ref{fig:scheme2}. We derive ML-AMP by first writing the belief
propagation equations \cite{yedidia2003understanding} for this graphical
model. As every factor relates to many variables and
every variable is contained in many factors, we use the central limit
theorem to keep track of only the means and variances of the
messages. Furthermore, we express the iterations in terms of node-variables
instead of messages, giving rise to the so called Onsager reaction
terms. This derivation was presented for the case of single-layer
generalized linear estimation in
e.g. \cite{rangan_generalized_2011,krzakala_probabilistic_2012}.
The resulting multi-layer-AMP (ML-AMP) algorithm 
then iteratively updates estimators of the means
$\bm{\hat h}^{(\ell)}$ and of the variances
$\bm{\sigma}^{(\ell)}, \ell=1,\dots, L$ of the auxiliary variables
$\bm{h}^{(\ell)}, \ell=1,\dots, L-1$ and of the signal $\bm{x}$. For
simplicity of the multi-level notation we denote
$\bm{\hat x} = \bm{\hat h}^{(L)}$, and similarly for the variance.

We first write the ML-AMP update equations for an intermediate layer
$\ell$ and then specify how they change for the very first ($\ell\!=\!1$)
and the very last ($\ell\!=\!L$) layers. At each layer $1 \!\ge\! \ell\! \ge\!
L$, there are two vectors, $\bm{V}^{(\ell)}\in {\mathbb
  R}^{n_{\ell-1}}$ and $\bm{\omega}^{(\ell)}\in {\mathbb
  R}^{n_{\ell-1}}$, associated with the factor nodes, and two vectors,
$\bm{A}^{(\ell)}\in
{\mathbb R}^{n_{l}}$ and $\bm{B}^{(\ell)}\in {\mathbb
  R}^{n_{l}}$, associated with the variable nodes. Their update reads
\begin{equation}
    \begin{aligned}
    V^{(\ell)}_{\mu} (t) &= \sum_i {\big[ W_{\mu i}^{(\ell)}}\big]^2 \sigma^{(\ell)}_{i}(t), \\
    \omega^{(\ell)}_{\mu} (t) &= \sum_i W_{\mu i}^{(\ell)} {\hat h}^{(\ell)}_{i}
    (t)- V^{(\ell)}_{\mu} (t) g^{(\ell)}_{\mu} (t-1), \\
    A^{(\ell)}_{i} (t) &= -\sum_\mu \big[{W_{\mu i}^{(\ell)}}\big]^2
    \, \partial_\omega g^{(\ell)}_{\mu} (t), \\
    B^{(\ell)}_{i} (t)  &= \sum_\mu W_{\mu i}^{(\ell)}
    g^{(\ell)}_{\mu}(t) + A^{(\ell)}_{i} (t) {\hat h}^{(\ell)}_{i}(t).
  \end{aligned}
  \label{eq:ML-AMP}
\end{equation}
To define functions 
$g^{(\ell)}_{\mu} (t)$, $\partial_\omega g^{(\ell)}_{\mu} (t)$ and to state how to use eqs.~(\ref{eq:ML-AMP}) to update
the estimators ${\hat h}^{(\ell)}_{i}(t)$ and variances
${\sigma}^{(\ell)}_{i}(t)$, we need to define an auxiliary function
${\cal Z}^{(\ell)}$
for $2 \ge l  \ge L$ as
\begin{equation}
    \begin{aligned}
    & {\cal Z}^{(\ell)} (A^{(\ell-1)}, B ^{(\ell-1)}, V^{(\ell)},
    \omega^{(\ell)}) \equiv  \frac{1 }{\sqrt{2 \pi V^{(\ell)}}}  \\
    &\quad \int {\rm d}h \, {\rm d}z \, P^{(\ell)}_\text{out}(h | z) \,
    e^{-\frac{1}{2} A^{(\ell-1)} h^2 + B^{(\ell-1)} h} \, e^{-\frac{(z -
    \omega^{(\ell)})^2}{2 V^{(\ell)}} } \, . \nonumber
    \end{aligned}
\end{equation}
With this definition, the estimators of marginal mean of the auxiliary
variables 
$\hat{h}^{(\ell)}_i, 1 \ge \ell \ge L-1$, and the
quantity $g^{(\ell)}_{\mu},  2 \ge \ell \ge L$, is computed as
{\small
\begin{equation}
    \begin{aligned}
    g^{(\ell)}_\mu(t) &= {\partial_\omega} \log
       {\cal Z}^{(\ell)}  (A^{(\ell-1)}_\mu , B_\mu^{(\ell-1)} ,
       V^{(\ell)}_\mu , \omega^{(\ell)}_\mu ), \\
    \hat{h}^{(\ell)}_i(t+1)  &= {\partial_B} \log {\cal Z}^{(\ell+1)}  (A^{(\ell)}_i ,
       B^{(\ell)}_i , V^{(\ell + 1)}_i , \omega^{(\ell + 1)}_i ),
    \end{aligned}   
    \label{eq:gandh}
\end{equation}
}%
where the quantities on
the right hand side of (\ref{eq:gandh}) are evaluated at time index $t$. 

The output function $g^{(1)}_\mu$ in the first layer is obtained as in the
standard AMP algorithm for generalized linear estimation:
\begin{equation}
{\cal Z}^{(1)}  (y, V^{(1)}, \omega^{(1)}) = \int {\rm d}z \, P^{(1)}_\text{out}(y | z) \, \frac{e^{-\frac{(z -
                \omega^{(1)})^2}{2 V^{(1)}}}}{\sqrt{2 \pi V^{(1)}}} \ ,
\end{equation}
\begin{equation}
g^{(1)}_\mu(t) = {\partial_\omega} \log
{\cal Z}^{(1)}  (y_\mu, V^{(1)}_\mu(t) , \omega^{(1)}_\mu(t) )\ .
 \end{equation}
In the last layer, the estimator
$\hat{h}^{(L)}_i= {\hat x}_i$ is 
obtained from
\begin{equation}
 \begin{aligned}
 & {\cal Z}^{(L+1)}  (A^{(L)}, B^{(L)}) = \int {\rm d}h \, P_X(h) \,
  e^{-\frac{1}{2} A^{(L)} h^2 + B^{(L)} h}, \, \\
&\hat{h}^{(L)}_i(t+1)  = {\partial_B} \log {\cal Z}^{(L+1)}  (A^{(L)}_i(t) ,
B^{(L)}_i(t) ).
    \end{aligned}   
\end{equation}
Finally, the expressions  $\partial_\omega g_\mu^{(\ell)} $ and
$\sigma_i^{(\ell)}$ in (\ref{eq:ML-AMP}) are defined as:
$\partial_\omega g^{(\ell)}_\mu(t) = \partial^2_\omega \log{{\cal
    Z}^{(\ell)}(t)}$ and $\sigma^{(\ell)}_i(t+1) = \partial^2_B \log{{\cal Z}^{(\ell+1)}(t)}$.

Note that the ML-AMP is closely related to AMP for generalized linear
estimation \cite{rangan_generalized_2011}. The form of ML-AMP is the
one we would obtain if
we treated each layer separately, while
defining an effective prior $P^{\rm eff}_X$ depending on the variables
of the next
layer, and an effective output channel $P^{\rm
eff}_\text{out}$ depending on the variables of the preceding layer:
{\small
\begin{equation}
    \begin{aligned}
      \kern-0.5em P^\text{eff}_X (h^{(\ell)} | V^{(\ell + 1)} \!, \omega^{(\ell +
          1)} \!) & \! = \! \! \! 
            \int \! \! {\rm d}z \, P^{(\ell+1)}_\text{out} (h^{(\ell)} | z) \frac{e^{-\frac{(z -
            \omega^{(\ell + 1)})^2}{2 V^{(\ell + 1)}}}}{\sqrt{2 \pi V^{(\ell + 1)}}}, \\
      \kern-2em  P^\text{eff}_\text{out} (z^{(\ell)} | A^{(\ell-1)} \!,
        B^{(\ell-1)} \!) & \!=  \!  \! \! \int  \! \!  {\rm d}h \,
        P_\text{out}^{(\ell)}(h | z^{(\ell) \!}) \,
        e^{-\frac{1}{2} A^{(\ell-1)} h^2 + B^{(\ell-1)} h}.\nonumber
    \end{aligned}
\end{equation}
}%

\begin{figure}
    \algnewcommand\algorithmicto{\textbf{to}}
    \algrenewtext{For}[3]%
    {\algorithmicfor\ $#1 \gets #2$ \algorithmicto\ $#3$ \algorithmicdo}
    \begin{algorithmic}[1]
    \Procedure{ML-AMP}{}
        \State initialize $g_\mu^{(\ell)} = 0 \; \forall (\mu, \ell)$
        \State initialize $h_i^{(\ell)} = 0, \sigma_i^{(\ell)} = 1 \; \forall (i, \ell)$
        \For{t}{1}{t_\text{max}}
        \For{\ell}{1}{L}
            \State compute $V_\mu^{(\ell)}, \omega_\mu^{(\ell)} \; \forall \mu$ using (\ref{eq:ML-AMP})
            \State compute $g_\mu^{(\ell)}, \partial_\omega g_\mu^{(\ell)} \; \forall \mu$ using (\ref{eq:gandh})
            \State compute $A_i^{(\ell)}, B_i^{(\ell)} \; \forall i$ using (\ref{eq:ML-AMP})
        \EndFor
        \State compute $\hat{h}_i^{(\ell)}, \sigma_i^{(\ell)} \; \forall (i, \ell)$ using (\ref{eq:gandh})
        \EndFor
    \EndProcedure
    \end{algorithmic}
\end{figure}


{\bf State evolution:} A very useful property of AMP, compared to other algorithms
commonly used for estimating marginals of 
posterior distributions such as (\ref{posterior}), is that its performance can
be traced analytically in the thermodynamic limit using the so-called {\it
  state evolution} (SE)
\cite{bayati2011dynamics,rangan_generalized_2011}. This is a version
of the density evolution \cite{mezard2009information} for dense
graphs. It is known as the
cavity method in statistical physics \cite{mezard2009information}, where it is in general
non-rigorous.

Specifically, the cavity method implies that at each layer~$\ell$, the
overlap between the ground true $\bm{h}^{(\ell)}$ and its estimate
$\bm {\hat h}^{(\ell)}(t)$ provided by the ML-AMP algorithm at
iteration $t$ concentrates around an ``overlap'' $m^{(\ell)}(t)$
\begin{equation}
{m}^{(\ell)}(t)  \stackrel{n_\ell \to \infty}{=} \frac{1}{n_\ell} \sum_{i =
  1}^{n_\ell} h^{(\ell)}_i {\hat h}^{(\ell)}_i(t)\, .\label{overlap}
\end{equation}
In the case of the  Bayes-optimal
inference of $\bm{x}$, the state evolution
for the ML-AMP algorithm is written also
in terms of a parameter
$\hat{m}^{(\ell)}$, defined as the value around which
the components of $\bm{A}^{(\ell)}$ concentrate. 

For the intermediate layers, $2 \ge \ell\ge L $, the SE reads
{\small
\begin{equation}
\begin{aligned} 
    &\hat{m}^{(\ell)}(t) = -\alpha_{\ell} \; \mathbb{E}^{(\ell)}_{h,
      z, b, w} 
\partial_\omega g^{(\ell)} \big( \hat{m}^{(\ell-1)} (t),
        b,\rho_{\ell} - m^{(\ell)}(t) , w\big), \label{eq:se2}
\\
&m^{(\ell-1)}(t+1) = \mathbb{E}^{(\ell)}_{h, z, b, w} 
 h\,  \hat h^{(\ell-1)} (\hat{m}^{(\ell-1)}(t), b, \rho_{\ell} -
        m^{(\ell)}(t), w), 
\end{aligned}
\end{equation}
}%
where the scalar functions $\partial_\omega g^{(\ell)}$ and
$\hat h^{(\ell)}$ are defined in eq.~(\ref{eq:gandh}). The quantity
$\rho_{\ell}$ is given by the second moment of the components of the
vector $\bm{h}^{(\ell)}$. It can be computed from the knowledge of
$P_X$ and $P^{(k)}_\text{out} $ for $k = \ell, \dots, L$.  The
expectations are taken over the joint distribution
\begin{equation}
\begin{aligned}
    P^{(\ell)} (h, z, b, w) &= P^{(\ell)}_\text{out} (h | z) \,
    \mathcal{N}(z; w, \rho_{\ell} - m^{(\ell)}) \, \times \\
    &\kern-1em \mathcal{N} (b; \hat{m}^{(\ell - 1)} h, \hat{m}^{(\ell - 1)}) \,
    \mathcal{N} (w; 0, m^{(\ell)}). \label{SE_prob}
\end{aligned}
\end{equation}
In the above distribution, $h,z,b,w$ are random variables representing
respectively: the
ground truth hidden variables at layer $(\ell-1)$, the components of the estimator of $W^{(\ell)} \bm{h}^{(\ell)}$,
the components of $\bm{B}^{(\ell-1)}$, and the components of
$\bm{\omega}^{(\ell)}$. Note that the probability distribution
(\ref{SE_prob}) is similar to the one appearing in the single-layer
G-AMP algorithm of \cite{rangan_generalized_2011} with the exception
that, in $P^{(\ell)}_{\rm out}(h|z)$,
 a known measurement
is replaced by the distribution of the hidden variable $h^{(\ell-1)}$
at the previous layer. This makes the multi-layer SE quite
intuitive for readers well familiar with the SE for G-AMP. 

At the first (leftmost) and last (rightmost) layers, the SE order
parameters are given by the same fixed point equations as in the
single-layer G-AMP setting, that is 
\begin{equation}
    \begin{aligned}
       & \hat{m}^{(1)}(t) = -\alpha_1 \; \mathbb{E}_{y, z, w} \, \partial_\omega g^{(1)} (y, \rho_1 - m^{(1)}(t),w), \\
       & m^{(L)}(t+1) = \mathbb{E}_{x, b} \, x \, \hat{h}^{{(L)}} (\hat{m}^{(L)}(t), b),
    \end{aligned}
    \label{eq:se1}
\end{equation}
where expectations are taken over $P(y, z, w) = P^{(1)}_\text{out} (y | z) \,
\mathcal{N} (w; 0, m^{(1)}) \, \mathcal{N} (z; w, \rho_1 - m^{(1)})$ and $P(x, b) = P_X
(x) \, \mathcal{N} (b; \hat{m}^{(L)} x, \hat{m}^{(L)})$ respectively. Thus, if $L = 1$, the
state evolution equations of ML-AMP reduce to that of the standard G-AMP
algorithm.

The state evolution are iterative equations. We initialize
$m^{(\ell)}(t=0)$ close to zero (or otherwise, corresponding to the
initialization of the ML-AMP algorithm), then compute
$\hat m^{(\ell)}(t=0)$ for all layers. We then compute $m$ for the next
time step for all layers, then $\hat m$ at the same time step and all
the layers, etc. Finally to obtain the mean-squared error on $h^{(\ell)}$
from the state evolution, we evaluate
${\rm MSE}^{(\ell)} = \rho_{\ell} - m^{(\ell)}$.

\section{Free energy and phase transitions}
\label{free_energy}
We define the free energy of the posterior 
(\ref{posterior}) as 
\begin{equation}
      \phi = -\lim_{\{n_\ell\}\to \infty} \frac{1}{n_L}\log{Z(\bm{y})}
      \ ,
      \label{free_def}
\end{equation}
where $\lim_{\{n_\ell\}\to \infty}$ denotes the thermodynamic limit.
This free energy is self-averaging, meaning that the above limit is
with high probability independent of the realization of the random
variable $\bm{y}$, it only depends on the parameters of the model
$\alpha_\ell$, $L$, $P_X$, and $P_\text{out}^{(\ell)}$.  A computation
analogous to the one that leads from belief propagation to the ML-AMP
algorithm and its SE can be used to rewrite the Bethe free energy
\cite{yedidia2003understanding} into a single instance free energy
evaluated using the fixed points of the ML-AMP algorithm. Using
averaging analogous to the one of state evolution one then rewrites
this into the so-called replica symmetric free energy
\begin{equation}
    \begin{aligned}
        &\phi_{\rm RS}(\bm{m}, \bm{\hat{m}}) = \frac{1}{2} \sum_{\ell = 1}^L
            \tilde{\alpha}_\ell m^{(\ell)} \hat{m}^{(\ell)} - \tilde{\alpha}_L
            \mathcal{I}^{(L + 1)} (\hat{m}^{(L)}) \\
        & - \sum_{\ell = 2}^{L} \tilde{\alpha}_{\ell - 1} \mathcal{I}^{(\ell)}
            (m^{(\ell)}, \hat{m}^{(\ell - 1)}) - \tilde{\alpha}_0 \mathcal{I}^{(1)}
            (m^{(1)}).
    \end{aligned}
    \label{eq:freeenergy}
\end{equation}
with $\tilde{\alpha}_\ell = n_\ell / n_L$, and
{\small
\begin{equation}
    \begin{aligned}
        &\mathcal{I}^{(L + 1)} (\hat{m}^{(L)}) = \mathbb{E}_{x, b} \, \log \mathcal{Z}^{(L + 1)} ( \hat{m}^{(L)}, b ), \\
        &\mathcal{I}^{(\ell)} (m^{(\ell)}, \hat{m}^{(\ell - 1)}) = \mathbb{E}_{h, z, b, w}^{(\ell)} \log \mathcal{Z}^{(\ell)} (\hat{m}^{(\ell - 1)}, b,\rho_{\ell} - m^{({\ell})},w), \\
        &\mathcal{I}^{(1)} (m^{(1)}) = \mathbb{E}_{y, z, w} \, \log
        \mathcal{Z}^{(1)} ( y, \rho_1 - m^{(1)} , w).
    \end{aligned} \nonumber
\end{equation}
}%
One can check, by computing derivatives of this free energy with
respect to $m_\ell$ and $\hat{m}_\ell$, that the stationary points of
$\phi_{\rm RS}(\bm{m}, \bm{\hat{m}}) $ are fixed points of the state
evolution equations (\ref{eq:se2}) and (\ref{eq:se1}). Let us now use
(\ref{eq:se2}) and (\ref{eq:se1}) to express $\bm{\hat{m}}$ in terms
of $\bm{m}$ and consider the free energy $\phi_{\rm RS}(\bm{m})$ only
as a function of the overlaps $\bm{m}$ (\ref{overlap}). Define
\begin{equation}
    \begin{aligned}
    \phi_{\rm RS} &\equiv \min_{\bm{m}} \, \phi_{\rm RS}(\bm{m}), \\
    \bm{m}_{\rm IT} &\equiv \operatorname*{argmin}_{\bm{m}} \, \phi_{\rm RS}(\bm{m}).
        \label{mIT}
    \end{aligned}
\end{equation}
In the setting of Bayes-optimal inference, the replica symmetric free
energy $\phi_{\rm RS}$ is equal to the free energy
(\ref{free_def}). At the same time the minimum mean squared error of
the Bayes-optimal estimation of $\bm{x}$ is given by 
\begin{equation}
      {\rm MMSE} = \rho_L - {m^{(L)}_{\rm IT}} \, , 
\end{equation}
where $\bm{m}_{\rm IT}$ is defined in (\ref{mIT}) and
$\rho_L$ is the second moment of~$P_X$. This
has been recently proven for the single layer linear estimation
\cite{reeves2016replica,barbier2016mutual}, and based on statistical
physics arguments we conjecture it to be true also in the present problem. 

We divide the region of parameters of the present problem into three
phases.
If the MMSE is not low enough (defined in a way depending on the
application) we say that inference of the signal $\bm{x}$ is
information-theoretically \emph{impossible}. If the MMSE is
sufficiently low and the ML-AMP algorithm analyzed via state evolution
matches it, then we say inference is \emph{easy}. Finally, and most
interestingly, if MMSE is sufficiently low, but ML-AMP achieves worse
MSE we talk about a region of algorithmically \emph{hard} inference. 
Defining an algorithmically hard region by the performance of one given
algorithm might seem in general unjustified. However, in the case of
single layer generalized linear estimation we know of no other polynomial
algorithm that would outperform AMP in the thermodynamic limit for
random $W$. This leads us to the above definition of
the hard phase also for the present multi-layer problem.

\begin{figure*}
    \centering
    \includegraphics[height=37ex]{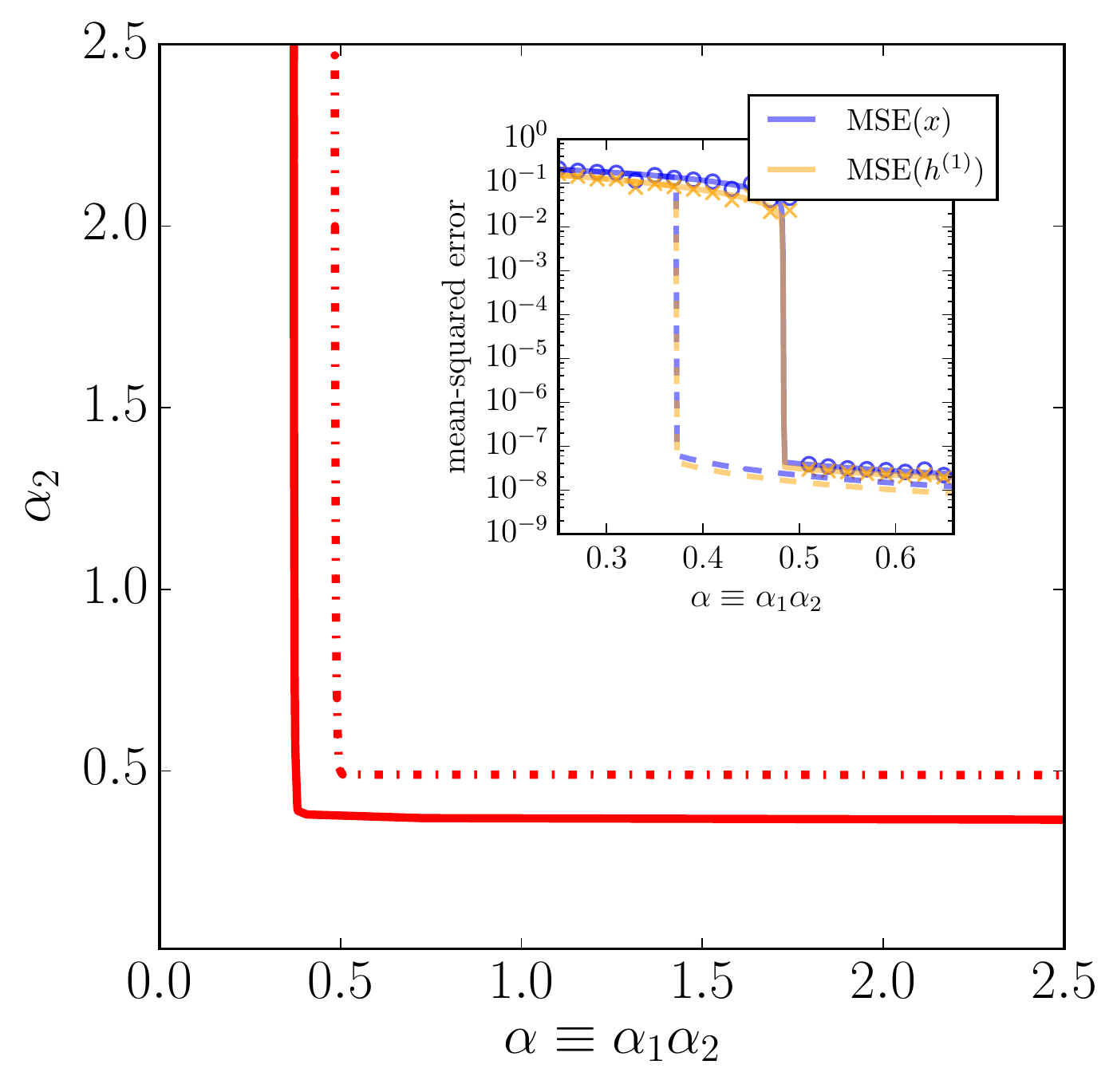}
    \includegraphics[height=37ex]{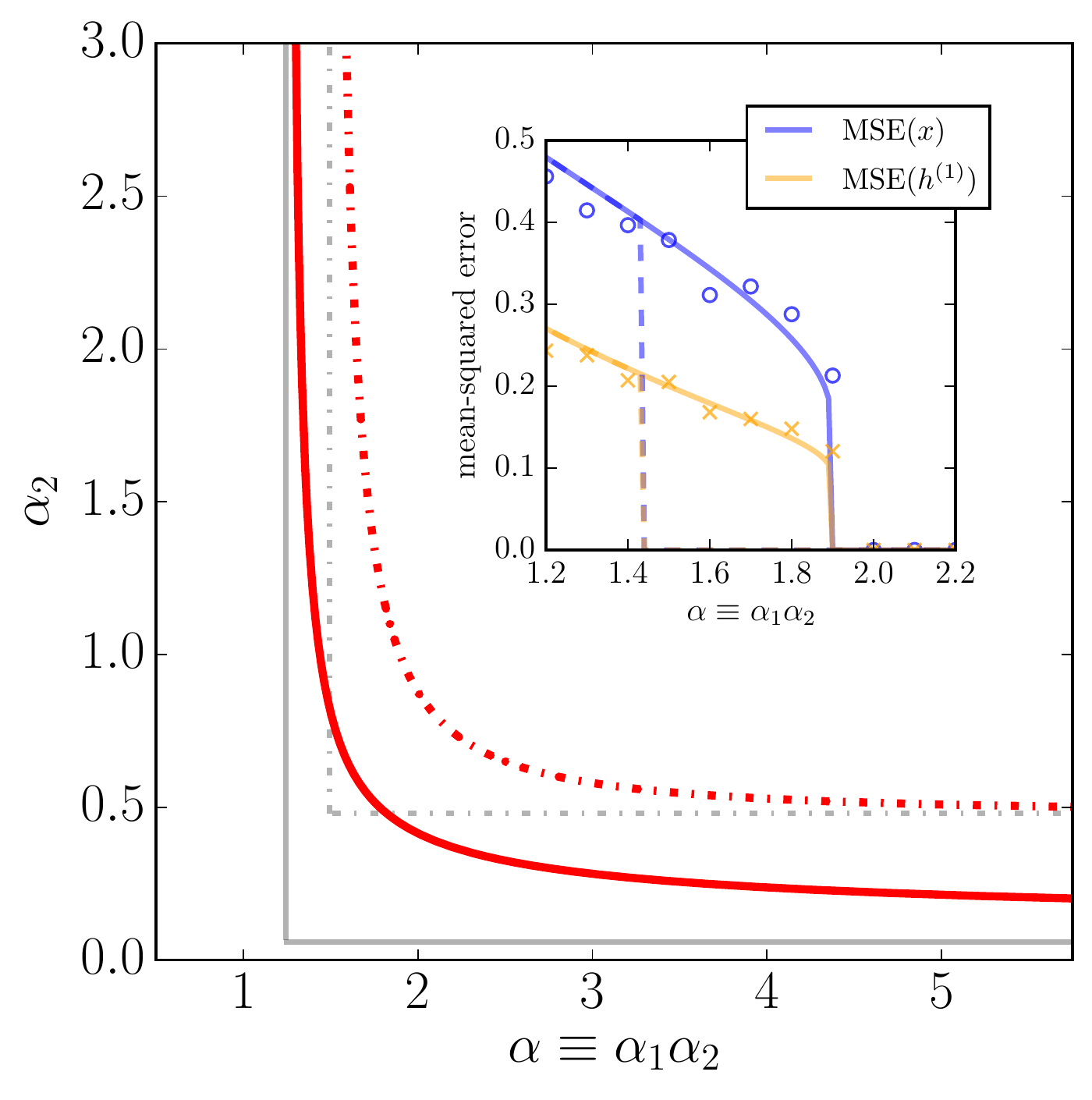}
    \includegraphics[height=37ex]{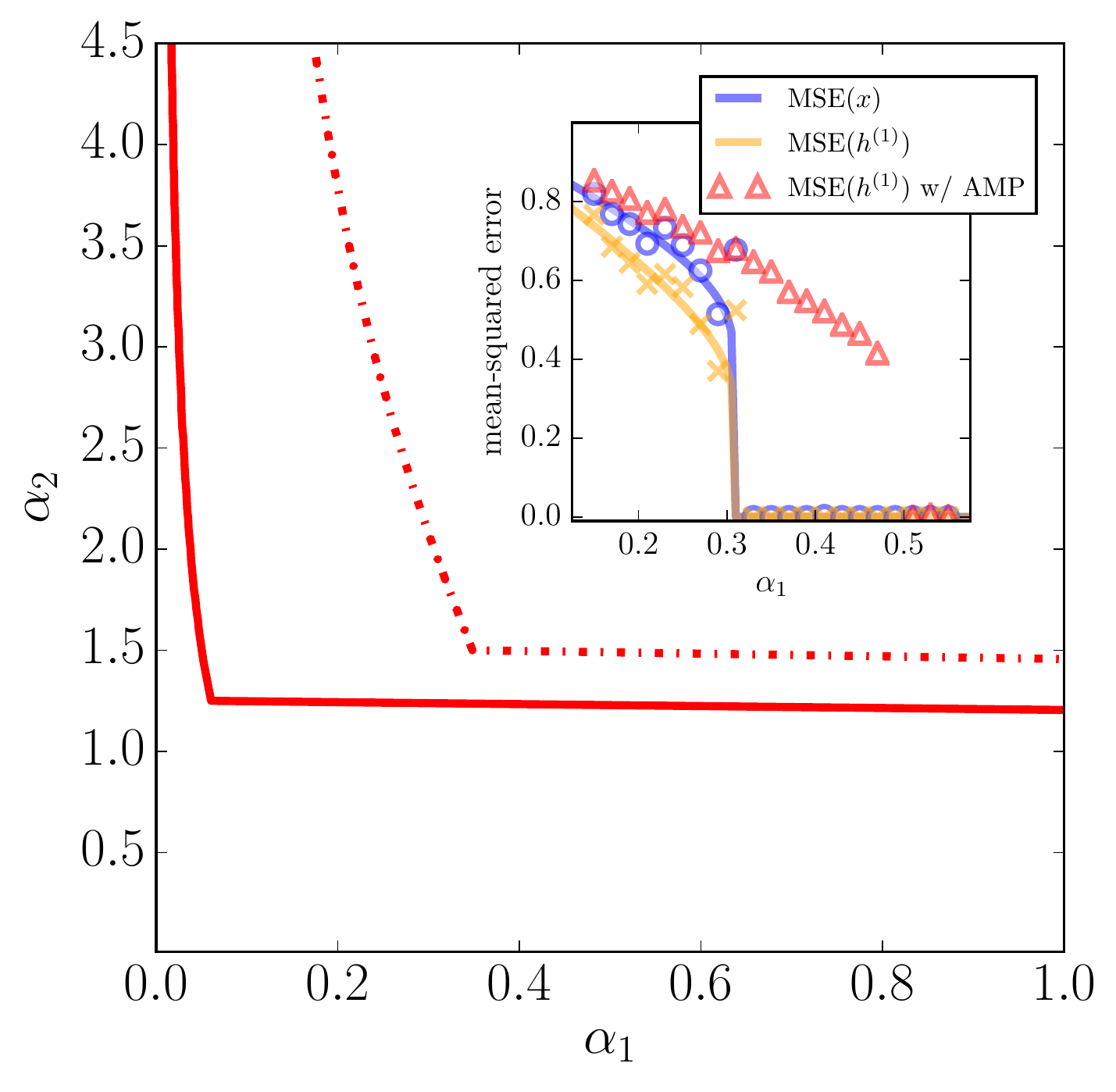}
    \caption{
    {\bf Main panels:} Phase diagram for sparse linear regression
    (left) and perceptron (center) with correlated data/patterns,
    defined by (\ref{eq:slr}) and
    (\ref{eq:perceptron}) respectively, and for the two-layer decoder (right),
    problem (\ref{eq:mixp}).
    Model parameters in SLR are $\rho = 0.3$, $\Delta_2 = 0$ and $\Delta_1 =
    10^{-8}$, and in the two-layer decoder $\Delta_1 = \Delta_2 = 10^{-8}$.
    The ML-AMP algorithm succeeds to reconstruct the signal with very
    small MSE above the dotted
    red line. Between the two lines reconstruction is
    information-theoretically possible, but ML-AMP does not achieve it.
    Below the full red line good reconstruction is impossible. The grey lines
    plotted for the perceptron is a comparison with the case of random
    patterns.  {\bf Insets:} Comparisons between the MSE predicted by the state
    evolution (lines) and that provided by ML-AMP on a single instance with
    $n_\ell = 2000$ (symbols), $\alpha_2 = 1.0$ for the left and center
    figures, and $\alpha_2 = 2.0$ for the right. Dashed lines indicate the
    MMSE. 
    Red triangles in the right inset compare to the performance of
    normal AMP in
    solving the decoder problem in the first layer assuming a binary
    i.i.d. prior on $h^{(1)}$, this works for $\alpha_1 \gtrsim 0.48$. 
    ML-AMP takes into account correlations in $h^{(1)}$ and
    performs better. 
    }
     \label{fig:adaptive}
\end{figure*}

\section{Results for selected problems}
We now focus on three examples of two-layer models and draw their
phase diagrams that indicate for which layer sizes (i.e. for which
values of $\alpha_\ell$) reconstruction of the signal is
information-theoretically possible. We also run the ML-AMP on single
instances sampled from the model and illustrate that the mean-squared
error it reaches after convergence matches the one predicted by the
state evolution.

{\bf Sparse linear regression using correlated data:} Among the
simplest non-trivial cases of the ML-GLM model is a two-layer analogue
of sparse linear regression (SLR) defined as
\begin{equation}
    \bm{y} = W^{(1)} (W^{(2)} \bm{x} + \mathcal{N} (0, \Delta_2)) +
        \mathcal{N} (0, \Delta_1).
    \label{eq:slr}
\end{equation}
where the vector $\bm{x}$ we seek to infer is sparse, $P_X (\bm{x}) =
\prod_{i = 1}^{n_2} [ \rho \, \mathcal{N} (x_i; 0, 1) + (1 - \rho) \,
\delta(x_i)]$, and $W^{(\ell)}_{\mu i} \sim \mathcal{N} (0, 1/n_\ell)$.

When $\Delta_2 = 0$ the model (\ref{eq:slr}) is equivalent to a
SLR with a structured data matrix $\Phi = W^{(1)}
W^{(2)}$, a problem previously studied by
\cite{0295-5075-76-6-1193,tulino_support_2013,kabashima_signal_2014,cakmak_s-amp:_2014,rangan_vector_2016}. Interestingly,
the state evolution equations (\ref{eq:se2}) have at their fixed point $\hat{m}^{(2)} = R \left(-(\rho_2 - m^{(2)})/\Delta_1\right)/\Delta_1$,
where $R$ gives the R-transform of $\Phi^T \Phi$ (see e.g.
\cite{tulino_random_2004}). Together with eq.~(\ref{eq:se1}) these are the
same equations obtained by adaptive approaches in
\cite{0295-5075-76-6-1193,tulino_support_2013,kabashima_signal_2014,cakmak_s-amp:_2014,rangan_vector_2016}. This
confirms that the adaptative methods are exact in the case of matrix
product, as was already noted for the Hopfield model~\cite{Mezard2016}.


In Fig. \ref{fig:adaptive} (left) we draw the phase diagram as a function of $\alpha \equiv \alpha_1 \alpha_2$ and
$\alpha_2$. Remarkably, in the noiseless case, the phase diagram of
the problem with structured matrix $\Phi$ is reduced to the statement
that both $\alpha$ and $\alpha_2$ need to be larger than the
corresponding threshold known for the usual compressed sensing problem
\cite{krzakala_probabilistic_2012}. 

For $\Delta_2 > 0$ this simple mapping does not hold and, as far as we
know, the ML-AMP and its SE analysis give new results.  We compared numerically
the performance of ML-AMP to the VAMP of
\cite{rangan_vector_2016}. Whereas for $\Delta_2=0$ the two algorithm
agree (within errorbars), for $\Delta_2>0$ the ML-AMP gives a
distinguishably lower MSE.

{\bf Perceptron learning with correlated patterns:} 
A lot of work  has been dedicated to learning and generalization in a
perceptron with binary weights \cite{gardner1989three,engel2001}, defined by:
\begin{equation}
    \bm{y} = \operatorname{sgn} (\Phi \bm{x})
    \label{eq:perceptron}
\end{equation}
with $\bm{x} \sim \prod_{i = 1}^{n_2} [\frac{1}{2} \delta (x_i - 1) +
\frac{1}{2} \delta(x_i + 1)]$.
These works focused on random patterns where the elements of $\Phi$
are iid.

 It was recently argued that learning and generalization of combinatorially structured
patterns, defined as  $\Phi = W^{(1)} W^{(2)}$, is best studied using multilayer
networks and presents major
differences \cite{Mezard2016}.
Our analysis of this case in Fig.~\ref{fig:adaptive}
(center) quantifies how many extra samples are needed so that a binary
perceptron is able to correctly classify combinatorially correlated patterns with
respect to random ones.

{\bf Two-layer decoder: }The most exciting potential
applications for the present results perhaps lie in the realm of deep neural
networks where models such as ML-GLM with learned weights $W^{(\ell)}$
are used. A crucial ingredient of such neural networks are non-linear
activation functions present among the layers. These activation
functions can be seen as noisy channels, e.g. in the context of
the decoder-side of an auto-encoder with known weights $W^{(\ell)}$: one
is interested in how well a vector of latent variables  can be
reconstructed when $y$ is observed at the output. In
Fig.~\ref{fig:adaptive} (right) we draw a phase diagram for the following example
\begin{equation}
    \bm{y} = W^{(1)} \operatorname{sgn} (W^{(2)} \bm{x} +
        \mathcal{N} (0, \Delta_2)) + \mathcal{N} (0, \Delta_1), 
    \label{eq:mixp}
\end{equation}
with
$\bm{x} \sim \prod_{i = 1}^{n_2} [\frac{1}{2} \delta (x_i - 1) +
\frac{1}{2} \delta(x_i + 1)]$.
Our results illustrate that the ML-AMP algorithm provides better
results than a layer-wise estimation with ordinary AMP, because it
takes into account correctly the correlations among the hidden variables.

\section*{Acknowledgment}
This work has been supported by the ERC under the European Union's FP7
Grant Agreement 307087-SPARCS.



\bibliographystyle{IEEEtran}
%



\bibliography{zotero}

\end{document}